\documentclass[12pt]{iopart}
\usepackage{iopams}  
\usepackage{graphics, color}
\usepackage[sort]{cite}

\begin{document}

\title[Lai et al., Quenching NV center photoluminescence with infrared pulsed laser]{Quenching nitrogen-vacancy center photoluminescence with infrared pulsed laser}

\author{N D Lai$^{1,4}$, O Faklaris$^{1,4,5}$, D Zheng$^{1}$, V Jacques$^{1,2}$, H-C Chang$^{3}$,  J-F Roch$^{1,2}$, and F Treussart$^{1,2}$}

\address{$^{1}$Laboratoire de Photonique Quantique et Mol\'eculaire, UMR 8537 CNRS and \'Ecole Normale Sup\'erieure de Cachan, 94235 Cachan cedex, France.}

\address{$^{2}$Laboratoire Aim\'e Cotton, CNRS UPR 3321, Universit\'e Paris Sud and \'Ecole Normale Sup\'erieure de Cachan, 91405 Orsay, France.}

\address{$^{3}$Institute of Atomic and Molecular Sciences, Academia Sinica, Taipei 106, Taiwan.}

\medskip
\address{$^4$~These authors have equally contributed to this work.}
\address{$^5$~\emph{Present address}: ImagoSeine, Institut Jacques Monod, UMR 7592 CNRS and Universit\'e Paris Diderot, 75205 Paris cedex 13, France}

\ead{nlai@lpqm.ens-cachan.fr; francois.treussart@ens-cachan.fr}

\begin{abstract}
Diamond nanocrystals containing Nitrogen-Vacancy (NV) color centers have been used in recent years as fluorescent probes for near-field and cellular imaging.
In this work we report that an infrared (IR) pulsed excitation beam can quench the photoluminescence of NV color center in a diamond nanocrystal (size~$<50$~nm) with an extinction ratio as high as $\approx 90\%$. We attribute this effect to the heating of the nanocrystal consecutive to multi-photon absorption by the diamond matrix.
This quenching is reversible: the photoluminescence intensity goes back to its original value when the IR laser beam is turned off, with a typical response time of hundred picoseconds, allowing for a fast control of NV color center photoluminescence. 
We used this effect to achieve sub-diffraction limited imaging of fluorescent diamond nanocrystals on a coverglass. For that, as in Ground State Depletion super-resolution technique, we combined the green excitation laser beam with the control IR depleting one after shaping its intensity profile in a doughnut form, so that the emission comes only from the sub-wavelength size central part.
\end{abstract}


\submitto{\NJP}

\maketitle

\section{Introduction}

The nitrogen-vacancy (NV) color center in diamond has attracted a large interest in various fields over the past ten years, owing to its unique properties, such as a perfectly stable photoluminescence (PL) allowing reliable single photon production~\cite{Beveratos:2001ht}, and a ground state $S=1$ electron spin resonance that can be optically readout~\cite{Jelezko:2004jn} and can be used for single spin manipulation in the solid state~\cite{Wrachtrup:2011uz}.

When embedded in a diamond nanocrystal or diamond nanostructure, the NV color center is used as a probe of the magnetic field at the nanoscale~\cite{Balasubramanian:2008ga,Rondin:2012ku,Maletinsky:2012ge} and as a fluorescent label in bio-applications~\cite{Chang:2008ia,Faklaris:2008bm}. 
In this context the perfect photostability not only allows long-term tracking in live cells~\cite{Fang:2011gm}, but also the use of stimulated emission depletion (STED) super-resolution microscopy~\cite{Rittweger:2009cc,ceggeli:2009wea,Tzeng:2011do}, which can also be made spin-selective~\cite{Maurer:2010js}.

Recently, the effect of the environment temperature on NV$^-$ photoluminescence properties was investigated in diamond nanocrystals~\cite{Plakhotnik:2010cg} and bulk material~\cite{Toyli:2012gl}. The PL properties of emitters in a solid matrix are known to depend on temperature~\cite{g1994luminescent}, with the common feature of spectral narrowing at temperatures below $\approx 200~$K, but the response to heating at temperature above room temperature is more material dependent.
In the case of diamond, a decrease of the NV$^-$ center PL intensity was observed for temperature higher than 550~K: it is reduced by 70\% in 35~nm fluorescent diamond nanocrystals~\cite{Plakhotnik:2010cg} and by 80\% in bulk diamond~\cite{Toyli:2012gl}. In the later case, the effect was attributed to thermally activated nonradiative processes. Moreover, Plakhotnik and Chapman~\cite{Plakhotnik:2011fy} showed that high energy 532~nm wavelength pulsed laser excitation of NV-containing diamond nanocrystals is also capable of inducing a 60\% PL quenching attributed to diamond lattice temperature increase.

In this work we show that we get a similar decrease of the photoluminescence intensity up to 92\%, by focusing a strong pulsed infrared (IR) laser beam (at 1064~nm wavelength) on fluorescent diamond nanocrystals (fND) deposited on a glass coverslip. We investigated the time dependence of the effect with the delay between the IR and green (532~nm wavelength, used for PL excitation) pulses. We observed that the effect is weaker when the IR laser beam is focused on NV centers hosted in a bulk diamond sample. 
Our observations are consistent with an increase of the nanocrystal temperature above 550~K resulting from five-photon absorption of the IR beam by the diamond matrix followed by a heat dissipation in the surrounding environment much slower than in the case of bulk diamond which has a high thermal conductivity.

As an application of this PL quenching effect, we imaged single fNDs below the diffraction limit, by applying a similar strategy as in STED~\cite{Rittweger:2009cc} or Ground State Depletion (GSD)~\cite{Rittweger:2009gv} super-resolution microscopies.
The IR pulsed-laser beam is used to control NV center PL intensity like the absorption transition saturating beam in GSD. As a proof of concept, we demonstrated a 28\% reduction of the fluorescence spot width in super-resolution imaging of size $\approx30$~nm fNDs using the superposition of a doughnut-shaped IR-laser beam and the Airy spot of the fluorescence excitation beam.

This article is organized as follows. In Section~2, we present the experimental setup and the diamond samples used for the study. In Section~3, we report the modification of the fND photoluminescence under pulsed IR illumination and we discuss the mechanism that could be responsible for the PL quenching observed. Section 4 presents the application of such an IR excitation to super-resolution imaging of NV color centers in diamond nanocrystals. The last section is dedicated to the summary and prospects.

\section{Experimental setup and diamond sample preparation}

\begin{figure}[b]
\centerline{\resizebox{\columnwidth}{!}{\includegraphics{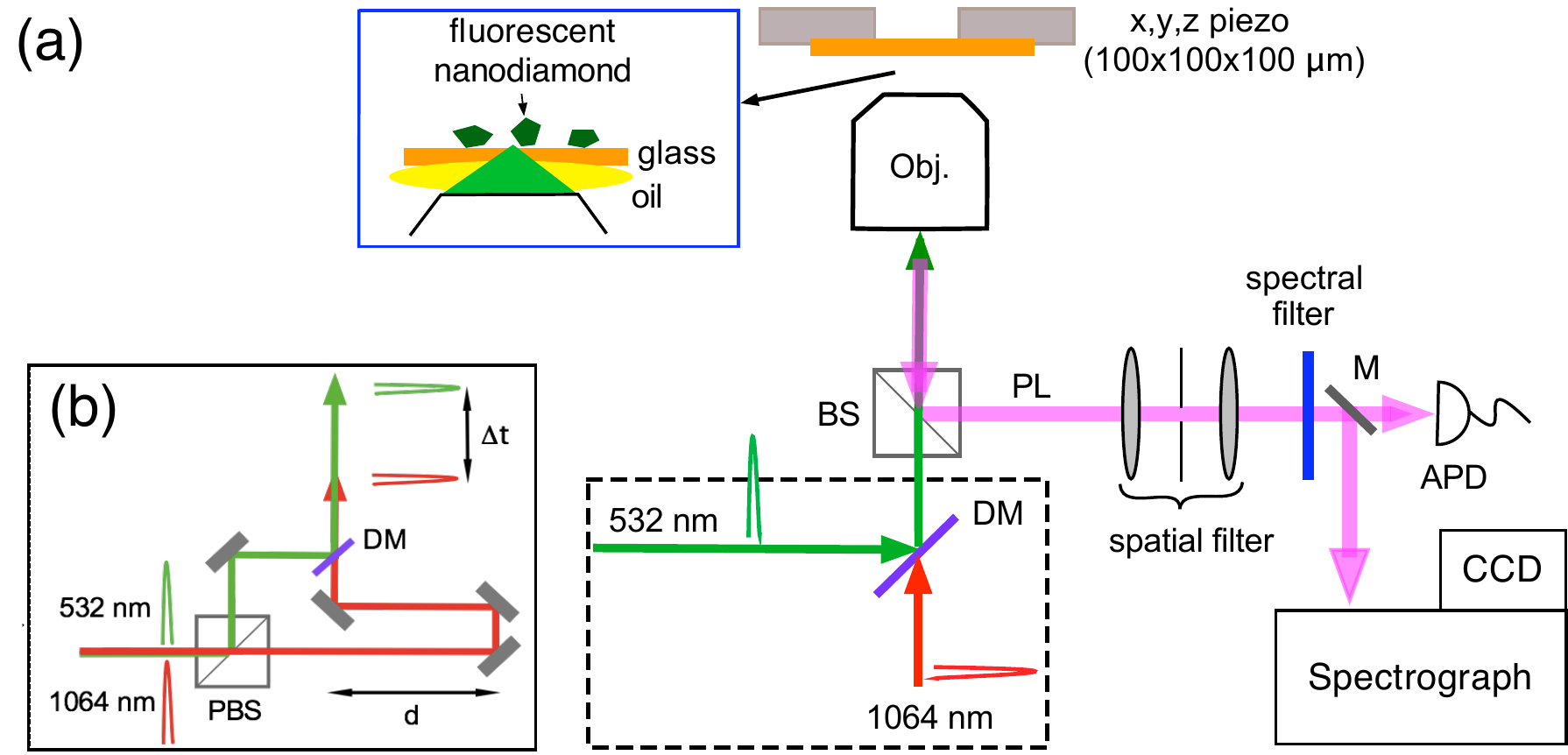}}}
\caption{Experimental setup consisting of a scanning confocal microscope with two input pulsed laser beams superimposed: the fundamental beam (1064~nm) of the home-made Nd-doped YVO$_4$ picosecond pulsed laser, and the frequency doubled one at 532~nm obtained after passing the IR beam through a 1.5~cm long KTP non-linear crystal. The 532~nm-wavelength laser beam serves to excite the NV center photoluminescence and , when added, the 1064~nm beam quenches the PL. (a) The 532~nm and 1064~nm beams are spatially superimposed with a dichroic mirror (DM) before being focused onto the sample via a high numerical aperture (oil immersion, $\times 60$, NA =1.4) objective (Obj). The focus point is raster scanned relative to the sample with nanometer resolution, using a 3D-piezoelectric translation stage. The PL is collected by the same microscope objective, sent towards the detection arm by a 50/50 broadband beam splitter cube (BS), and then spatially filtered (100~$\mu$m diameter pinhole). The NV center PL is spectrally filtered from the remaining 532~nm excitation laser by a long-pass filter (cut-off wavelength at 545~nm) and from 1064~nm illumination by a short-pass glass filter (Schott KG5, wavelength cut-off: 850~nm). The PL signal is then either integrated on a silicon avalanche photodiode-based single-photon counting detector (APD), or sent with a mirror (M) onto an imaging spectrograph equipped with a cooled CCD array detector, for spectral analysis. \emph{Inset} (blue box): zoom on the nanodiamond sample: 30~nm diamond nanocrystals spin coated on a coverglass.
(b) Configuration used, at the place of the dashed-line surrounded part of (a), to vary the time-delay $\Delta t$ between 1064~nm and 532~nm pulses, by varying the length $d$.}
\label{fig:setup}
\end{figure}

Optical excitation and detection of the NV color center photoluminescence (PL) is realized with a home-built scanning confocal microscope shown in figure~\ref{fig:setup}, using a 1.40 numerical aperture $\times60$ magnification plan apochromatic microscope objective (Nikon, Japan). The PL excitation laser source is the frequency-doubled beam (at 532~nm wavelength) of a home-made neodynium doped YVO$_4$ crystal (emission wavelength 1064~nm) pulsed laser operating in a mode-locked regime with a repetition rate of 4.8~MHz and a pulse duration of 16~ps~\cite{Papadopoulos:2003hv}. 
As shown later, when superimposed to the 532~nm beam and focused on diamond nanocrystals, the 1064~nm beam leads to the decrease of NV color center PL intensity.

We used high-pressure and high temperature (HPHT) synthetic diamond of type~1b, i.e. with $\approx 100$~ppm nitrogen impurity concentration. NV color centers were created in either diamond nanocrystals (nanodiamonds) or single crystal diamond plate. 
To create NV color centers in nanodiamonds, we started from commercially available diamond powder of 35~nm mean size (MSY grade from Microdiamant, Geneva, Switzerland) and used the same procedure as in Ref.\cite{Chang:2008ia}, consisting in an irradiation with a 40~keV energy He$^+$ beam to create the vacancies, followed by a thermal annealing at $800^{\circ}$C under vacuum leading to the migration of vacancies and their stabilization next to nitrogen impurities. After strong acid cleaning and dispersion in water, the fluorescent nanodiamonds were spincoated on a coverglass at a surface density sufficiently small to allow for single nanocrystal addressing with the confocal microscope.
We observed with the spectrograph that about half of the nanodiamonds displayed the characteristic spectrum of negatively-charged NV$^-$ color centers, and the other half of the population contains neutral charge NV$^{\circ}$~\cite{Rondin:2010dn}.

Regarding the bulk sample preparation, we irradiated a diamond single crystal plate (Element Six Ltd, UK) with 2~MeV  electron beam (dose $10^{13}$ electrons/cm$^2$) and then annealed it for 2~h at 850$^\circ$C in vacuum. This resulted in a density of about 200~NV color centers/$\mu$m$^3$ (mostly negatively charged)~\cite{Lai:2009gm}. 

\section{Quenching nanodiamond photoluminescence with 1064~nm pulsed laser illumination.}

\begin{figure}[b]
\centerline{\resizebox{\columnwidth}{!}{\includegraphics{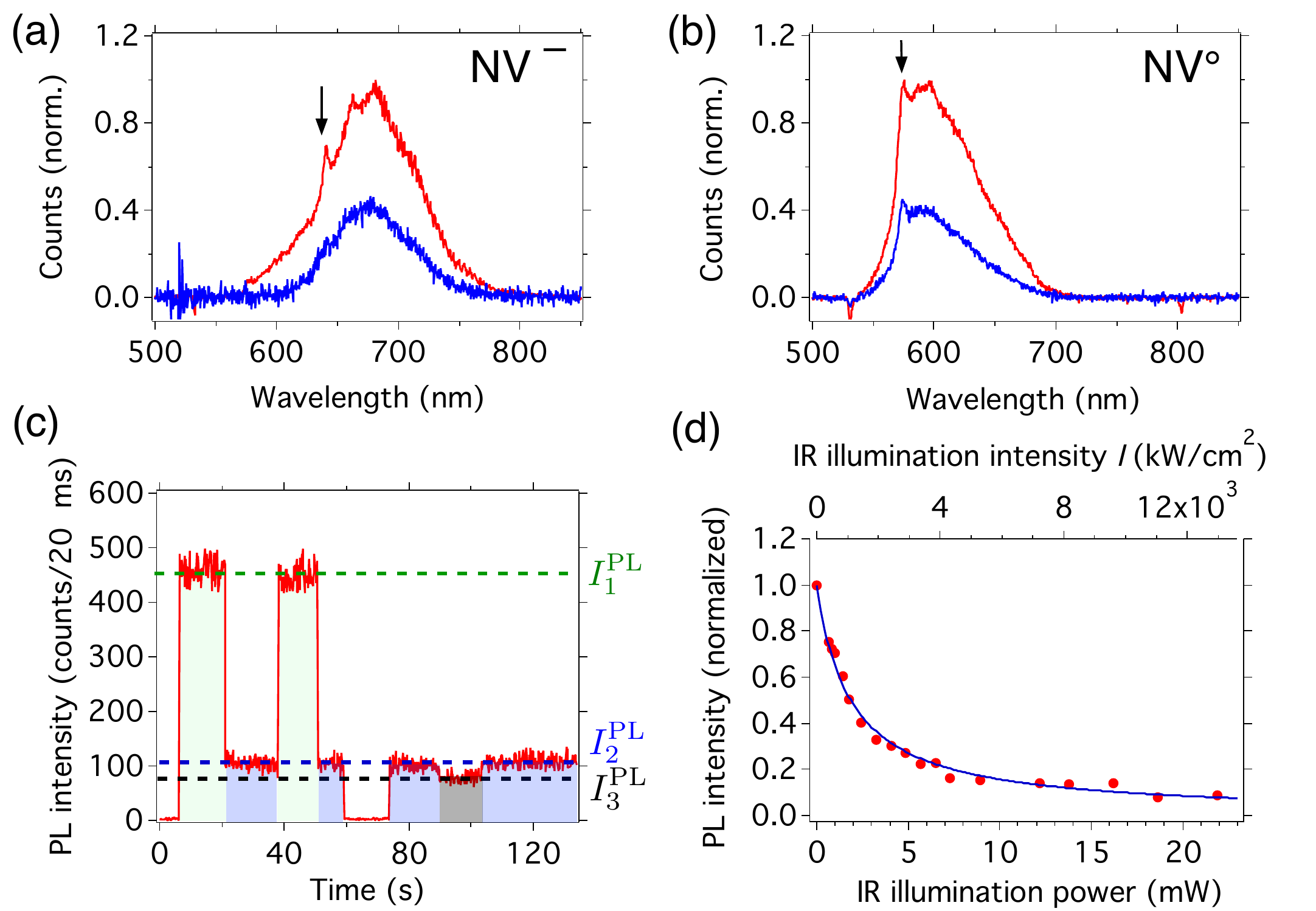}}}
\caption{Effect of the addition of the 1064~nm wavelength pulsed laser illumination on the photoluminescence of NV color centers in nanodiamonds excited by the frequency-doubled excitation laser beam (532~nm wavelength). (a) PL spectrum obtained without (red) or with (blue) the 1064~nm beam for NV$^-$-containing nanodiamonds. 
(b) PL spectrum for NV$^\circ$-containing nanodiamonds. 
For (a) and (b): the black arrows indicate the positions of the zero phonon lines (ZPL) characteristic of each NV center type (ZPL at 637~nm for NV$^-$ and 575~nm for NV$^\circ$). 
(c) PL intensity monitoring vs time. The green shaded area correspond to the situation where IR illumination is ON (intensity level $I_2$, blue dashed line). The blue shaded area corresponds to the IR illumination OFF (intensity level $I_1$, green dashed line). We observe that the PL quenching effect is reversible. The substrate background intensity with both lasers ON is indicated by grey shading (level $I_3$, black dashed line) and is obtained by moving both excitation spots $\approx1~\mu$m away from the nanodiamond. The detector dark counts is obtained by switching off the green excitation laser and is displayed as the non-shaded regions (see intensity associated to the 60--75~s time interval). The quenching efficiency reached is 92\%. Excitation power: 22~mW at 1064~nm and 0.15~mW at 532~nm. 
(d) For the same fND as in (c), dependence of background subtracted PL intensity vs IR pulse power (lower horizontal scale) and intensity $I$ in the focal plane (upper scale). The blue line is a fit by the saturation function $1/(1+I/I_{\rm sat})$ yielding a saturation intensity $I_{\rm sat}=1.18\pm0.03$~MW/cm$^2$.}
\label{fig:spectrum}
\end{figure}

\noindent 
Nanodiamonds containing NV color centers were localized on the substrate by recording their PL signal under the pulsed 532~nm excitation laser beam (average power of 0.15~mW) while raster scanning the sample. We checked with a time intensity correlation measurement setup (not shown on figure~\ref{fig:setup})~\cite{Beveratos:2001ht}, that most of the nanodiamonds  often contain more than one NV center.

\begin{figure}[b]
\centerline{\resizebox{\columnwidth}{!}{\includegraphics{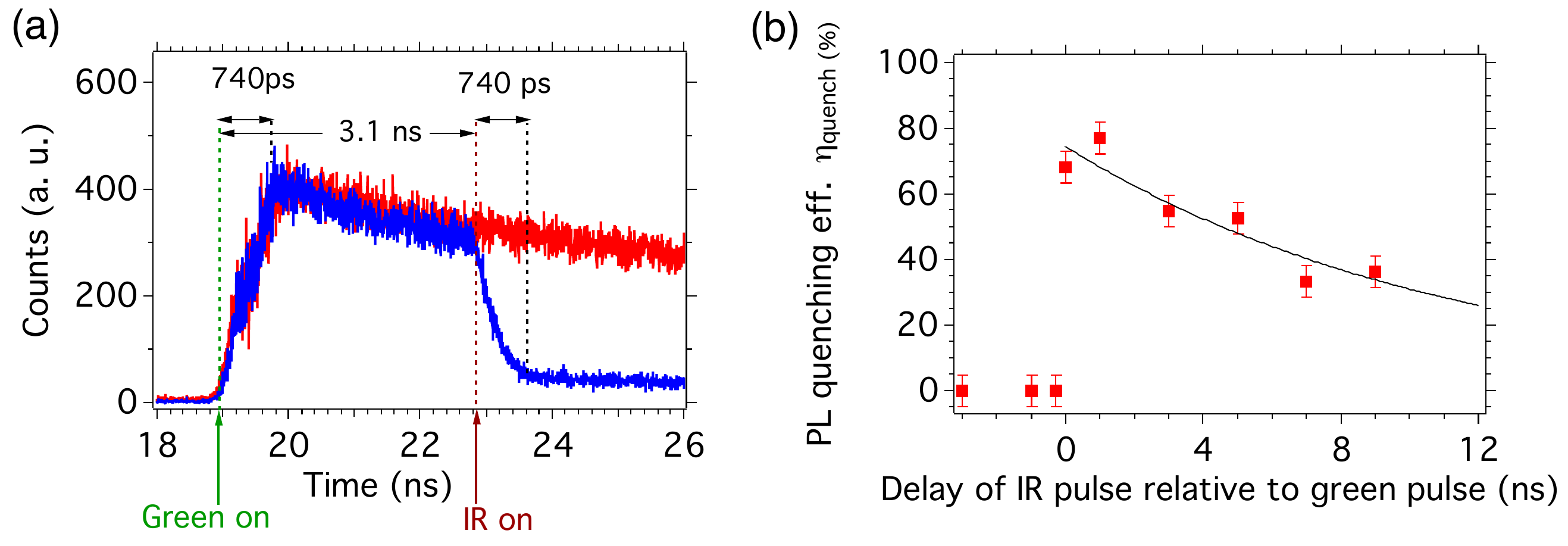}}}
\caption{Photoluminescence decay of NV color centers in nanodiamonds with or without the infrared beam, at various delays between IR (1064~nm) and green (532~nm) pulses. (a) The IR pulse is delayed by 3.1 ns with respect to the green pulse. The apparent response time of the PL quenching by the IR pulse is about 740~ps, which is indeed the response time of the detection system. (b) NV color center photoluminescence quenching as a function of the 532~nm -- 1064~nm pulses time-delay, with the convention of a positive delay in case of a 532~nm pulse ahead in time of the 1064~nm one. The line is an exponential fit used as a guide for the eyes, having a characteristic decay time of 11~ns related to NV center excited state lifetime, as expected. 
}
\label{fig:response_time}
\end{figure}

We then added the 1064~nm pulsed laser beam and observed a significant decrease of the PL intensity compared to 532~nm excitation alone, while we could have expected a small PL increase due to the small two-photon absorption cross-section of NV$^-$ center~\cite{Wee:2007dv}. This quenching is observed for both NV center charge state (figure~\ref{fig:spectrum}(a) and (b)). The effect is fully reversible: when the IR laser is switched off, the PL returns to its original intensity level (figure~\ref{fig:spectrum}(c)) on timescales shorter that the 20~ms integration time per point. This timescale is actually shorter than 1~ns as shown in time-resolved analysis of figure~\ref{fig:response_time}(b). 
We could observe it with a \emph{pulsed} 1064~nm beam only and not with a continuous wave laser at the same wavelength. The effect is also independent of the polarization state of both 532~nm and 1064~nm beams.

As a quantification parameter, we define the PL quenching efficiency $\eta_{\rm quench}=(I^{\rm PL}_1-I^{\rm PL}_2)/(I^{\rm PL}_1-I^{\rm PL}_3)$, where $I^{\rm PL}_i (i=1,2,3)$ are respectively the PL intensity of the fND with the 532~nm laser alone (1), with the addition of the 1064~nm beam (2), and with both lasers focalized 1~$\mu$m away from the fND (3, sample background), as displayed on figure~\ref{fig:spectrum}(c).
Figure~\ref{fig:spectrum}(d) shows that when we increase IR energy per pulse the NV center PL decreases, reaching a maximum quenching efficiency of 92\% around 20~mW mean IR power, as estimated after passing though the microscope objective (transmission of $\approx$ 80\% at 1064~nm). 
Note that the quenching efficiency varies significantly from one nanodiamond to the other. We studied it carefully for six fND, and observed maximal values from 54\% to 96\% with a median at 89\%. For some fND we had to use IR mean power up to 80 mW to reach this maximum. We attribute the quenching efficiency variations to differences in the highest temperature achieved for each nanocrystal, which is governed by parameters that may significantly vary from one fND to another (multi-photon absorption cross-section, heat transfer to the substrate...). Note that a further increase in IR energy per pulse above 80~mW usually produced sudden irreversible complete loss of the PL probably due to nanocrystal burning (i.e. its oxidation by ambient dioxygen leading to the production of carbon dioxide) consecutive to its heating above  $\approx 770$~K, considered as the ignition temperature for nanodiamond combustion~\cite{Xu:2002td}.

In order to determine the quenching response time, we did time-resolved fluorescence measurements using a time-to-amplitude converter (not shown on figure~\ref{fig:setup}). Figure~\ref{fig:response_time} (a) displays the decay of the  NV center photoluminescence with and without the IR beam, which is  superimposed to the green pulse after a delay set at 3.1~ns much shorter than the PL lifetime, which is of the order of 20~ns. 
We measured a quenching response time $\approx740$~ps, which is identical to the rising time accompanying the 532~nm excitation pulse, and appears to be equal to the jitter time of the avalanche photodiode photon counting modules. This indicates that the PL quenching effect happens at a characteristic time shorter than the instrumental response one, in the range of the pulse durations.

As mentioned above, the 1064~nm beam does not interact efficiently with the NV color center in its ground state~\cite{Wee:2007dv}. However, it has been shown that NV$^-$ color center has another radiative transition in the infrared range around 1046~nm corresponding to the transition between two metastable singlet states~\cite{Rogers:2008ip}, lying at lower energies than the triplet excited state. 
One possible mechanism explaining the decrease of the PL intensity under 1064~nm illumination could be the population of the metastable states from the excited state by intersystem crossing (ISC), followed by cycles within the metastable sub-level system, eventually leading to a ground state depletion. In that case we would expect that the quenching is more efficient with a larger green laser excitation power favoring ISC from the excited stated where the system spend more time. However, we did not observe such a dependance experimentally. Furthermore the quenching has a similar efficiency for NV$^\circ$ color center for which no IR transition was ever reported. We therefore ruled out a mechanism relying on metastable transitions.

We further investigated the effect with time-resolved PL measurement, we delayed the 1064~nm pulse with respect to the 532~nm pulse using an optical delay line (figure~\ref{fig:response_time}(b)), with a maximal delay limited to 9~ns by the 2.7~m optical table length. Figure~\ref{fig:response_time}(b) shows that the PL quenching is absent when the IR beam hits the nanodiamond ahead of the green excitation laser (negative delay), and that it is maximal at zero delay (within the 740~ps setup resolution) and then decreases with increasing positive delay.

These observations could results from two effects: (i) either the 1064~nm pulse is absorbed by the NV center into a higher energy level from which it decays in picosecond time range through non radiative channels, or (ii) the 1064~nm pulse is absorbed by the diamond matrix, and then transformed into heat, leading to a large temperature increase resulting in the change of the NV center photophysical parameters.
We ruled out the interpretation (i) because we observe exactly the same phenomenon whatever the charge state of the NV center is, while they have very different energy level structures.
Moreover, explanation (ii) is consistent with previous reports. The 1064~nm (1.16~eV) pulse can be absorbed by diamond through a five-photon process (total energy of 5.8~eV) leading to the creation of electron-hole pairs~\cite{Glinka:1999efa}, which provoke a temperature increase of the crystals upon their non-radiative recombination. We do not observe any of the radiative recombinations reported in~\cite{Glinka:1999efa}, probably because we address a single diamond nanocrystal and not a compressed pellet of nanocrystals with a large number of structural defects where electron-hole can 
also recombine radiatively.

Moreover, it was recently observed that the NV center quantum yield decreases at temperature larger than 550~K due to thermally activated non-radiative decay channels~\cite{Toyli:2012gl}. We therefore interpret NV center PL quenching as the consequences of the heating of the diamond nanocrystal due to IR absorption, provoking a very fast temperature increase above 550~K, that eventually leads to its burning when this temperature goes above $\approx 770$~K~\cite{Xu:2002td}, as mentioned above. 
The absence of quenching when the IR beam hits the ND less than 1~ns before the green laser excitation (figure~\ref{fig:response_time}(b)) indicates that the heat transfer allowing for the nanocrystal temperature to go back to a value at which the quantum yield is identical to the room temperature ones ($\approx 450~$K according to Ref.~\cite{Toyli:2012gl}), happens on sub-nanosecond timescale. In Appendix~A we provide estimates of the heat transfer characteristic times for either a purely convective or radiative process. These times are in the range of hundreds microseconds (radiative) to millisecond (convective) timescales, and are not consistent with our observations. Therefore, we believe that the heat transfer by conduction between the ND and the substrate is the dominant process.

This explanation of the PL quenching by nanodiamond heating is consistent with the fact that, in bulk diamond sample, at similar 1064~nm energy per pulse than for nanodiamonds, we observed that  $\eta_{\rm quench}\approx 20\%$ only (see Appendix~B figure~\ref{fig:bulk}(c)). This is probably due to the fast heat diffusion in bulk diamond owing to its very high thermal conductivity, preventing the temperature to reach the threshold of 550~K.
Despite the smaller increase of temperature, the diamond plate can be damaged by the IR laser pulse, as it shown on figure~\ref{fig:bulk}(d).  Actually, in the bulk the whole pulse energy is deposited onto the sample compared to only a small energy fraction in the case of nanodiamonds which have an absorption cross-section smaller than the focused beam area. 
Therefore, optical breakdown can take place in the diamond plate after the multi-photon absorption and the electron-hole pair formation~\cite{Zalloum:2010uv}. This breakdown effect consists of an avalanche ionization in which the seed electrons are driven by the laser field and then cause secondary collisional ionization~\cite{Joglekar:2004cr}. The energy gained by the laser-pulse driven seed electrons is likely to be much lower in the case of the confined nanometer scale diamond crystal than in the bulk, leading to a higher breakdown threshold for diamond nanocrystals.

\section{Application to imaging of fluorescent nanodiamonds below the diffraction limit}

We then applied the IR pulsed excitation quenching to super-resolution imaging of nanodiamonds containing NV color centers. We used a similar configuration as the one of STED~\cite{Rittweger:2009cc} or GSD~\cite{Rittweger:2009gv}.
In these two super-resolution techniques, an intense laser beam with a deep intensity minimum (doughnut shape beam), either depletes the excited state by stimulated emission (STED) or saturates the excitation transition leading to the population of a dark metastable state and a subsequent ground state depletion (GSD). The fluorescence is then probed with the Airy spot of a normally focused co-aligned second beam, and the signal comes only from the sub-diffraction limited doughnut center.

\begin{figure}[h]
\centerline{\resizebox{\columnwidth}{!}{
\includegraphics{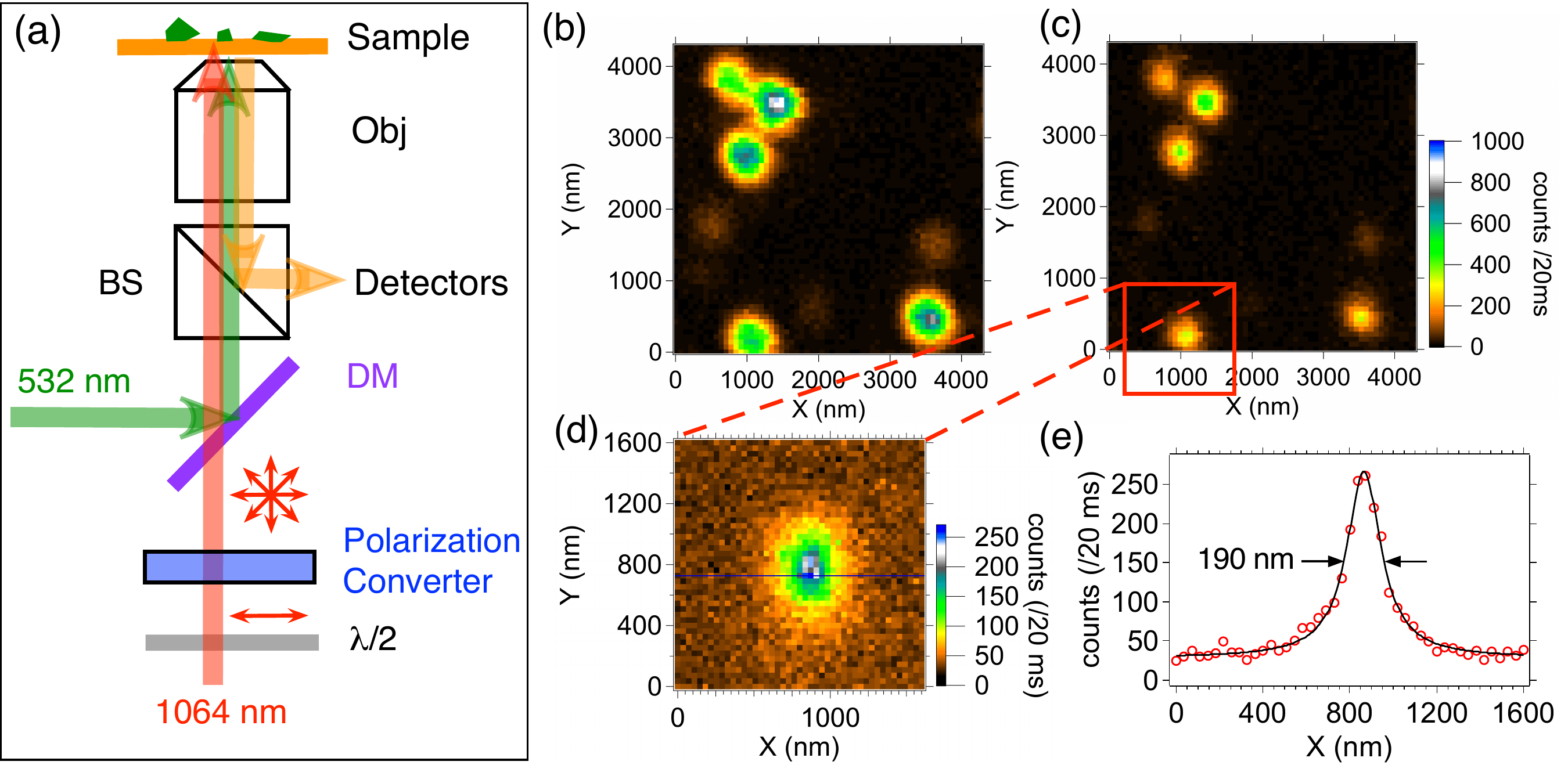}}}
\caption{Imaging fluorescent nanodiamonds below the diffraction limit using IR pulse quenching. (a) Experimental setup combining the Airy spot of the green laser (532~nm) and the IR laser beam (1064~nm) which linear polarization is transformed into a radial one with the use of a liquid-crystal-based polarization converter. Such a radially polarized beam has a doughnut-shaped intensity at the focus of the high numerical aperture microscope objective (Obj). BS: 50/50 broadband beamsplitter; DM: dichroic beamsplitter with high transmission at wavelength larger than 550~nm. 
(b) PL sample raster scan obtained with only the green excitation laser beam. (c) PL raster scan obtained with the superposition of the green and the IR doughnut spots (laser beam power: 0.15~mW at 532~nm and 60~mW at 1064~nm). (e) PL cross-section drawn along the blue horizontal line of spot (d). The line in (e) is a Lorentzian fit of the data (open circles) with a FWHM of $190\pm 6$~nm (below the diffraction limit).}
\label{fig:super-resolution}
\end{figure} 

Figure~\ref{fig:super-resolution}(a) displays the super-resolution imaging setup we used. The IR beam passes through a liquid-crystal based radial polarization converter (ARCoptix S.A., Switzerland). The focusing of such radially polarized cylindrical beam results in the doughnut shaped intensity at the focus of the microscope objective~\cite{Youngworth:2000hf}. The advantage of such a beam intensity shaping technique compared to the use of vortex phase mask~\cite{Rittweger:2009cc}, is that the liquid-crystal polarization converter covers a broad range of operating wavelength, including 1064~nm.
The non-modified 532~nm beam forming an Airy spot is superimposed to the doughnut in the sample plane in order to excite the NV center PL.
While raster scanning a diamond nanocrystal relative to the two laser spots, no modification of the PL intensity should be observed when the nanodiamond is located in the ``dark'' middle part of the IR doughnut beam. In this configuration, the ``bright'' doughnut ring hits the coverglass and not the nanodiamond, and thus does not lead to sufficient temperature increase and PL quenching. 
The nanodiamond PL spot shape results from the convolution of the particle shape and the doughnut beam, so that the ultimate imaging resolution, the one that should be obtained with an infinite power for the IR beam, is limited to the nanodiamond size~\cite{Greffet:2011tb}.

Figure~\ref{fig:super-resolution}(b) shows the PL rasterscan of a 30~nm diamond nanocrystal. When we add the IR pulsed beam to the 532~nm PL excitation beam (figure~\ref{fig:super-resolution}(c)), the spot full-width-at-half-maximum (FWHM) on an horizontal cross-section decreases down to $190\pm 6$~nm (figure~\ref{fig:super-resolution}(d) and (e)), corresponding to a reduction of  28\% compared to the diffraction limit 0.61/NA$\times \lambda=232~$nm, for NA=1.4 and $\lambda=532~$nm. The elliptical shape of the spot is the consequence of a non-symetrical doughnut illumination due to imperfections of the radial polarization converter.
The gain in resolution is moderate because of two factors: (i) the large size of the IR doughnut spot with respect to the size of the green excitation Airy spot, and (ii) the quenching efficiency. Considering the similarities between the PL non-radiative decay induced by heating and ground state depletion, we propose to use GSD formula provided in Ref.~\cite{Rittweger:2009gv} as an estimate of the theoretical FWHM resolution limit, in which contrary to GSD, the ultimate resolution is not limited by the doughnut contrast but rather by the quenching contrast $\varepsilon\equiv 1-\eta_{\rm quench}$: $\displaystyle\Delta r_{\rm theo}\approx\lambda_{\rm IR}/(\pi n)\sqrt{\varepsilon+I_{\rm sat}/I_{\rm m}}$, with $n$ being the refraction index of glass and $I_{\rm m}$  the crest intensity of the doughnut. The maximal quenching efficiencies were recorded in the range 90-96\% at the maximum power compatible with non-burning of diamond in air, but in the case of the fND of Fig.\ref{fig:super-resolution} the efficiency was only 54\%, yielding $\Delta r_{\rm theo}\approx 155$~nm, consistent with the 190~nm measured. With $\eta_{\rm quench}=92\%$ (see figure~\ref{fig:spectrum}(d)), we could expect $\Delta r_{\rm theo}\approx 65$~nm.

Note that we did not observe any resolution improvement for isolated NV center in the bulk diamond sample, as it was expected from the much smaller PL quenching obtained in the bulk.

\section{Conclusion}
We have shown that a high energy picosecond IR laser illumination at 1064~nm of NV center-containing diamond nanocrystals reversibly quenches the fluorescence with $\approx 90\%$ efficiency. This effect is maximal when the IR and fluorescence excitation pulse (at 532~nm) arrives simultaneously onto the nanodiamond, and it is absent when the 1064~nm pulse is ahead of the 532~nm one.
Considering recent reports on the reduction of NV center PL quantum yield upon heating, we proposed that the observed PL quenching  under IR pulsed illumination results from the heating of the nanodiamond, after the absorption of the 1064~nm beam by the diamond matrix through a five-photon process, up to temperature of about 600~K.

Such a high local temperature could be of interest for photothermal therapy applications, in which antibody targeting over-expressed proteins characteristic of the cells to be destroyed, are first grafted to the nanodiamond. Such a strategy has been employed over the past ten years with gold nanostructures relying on the coupling of incident light to plasmonic modes~\cite{Loo:2005iu}, up to 800~nm excitation wavelength. A related approach was even tested with detonation nanodiamonds, which can undergo a dramatic increase of size under intense illumination at 532~nm wavelength, provoking damages to the targeted cells~\cite{Chang:2008hj}.

As an application of our observations in nanophotonics, we combined a doughnut-shaped IR beam and a green non-modified ones to image diamond nanocrystals at  a resolution below the diffraction limit. We obtained a FWHM spot reduction of 28~\% compared to standard confocal microscopy, which was in this nanodiamond case limited by a low quenching efficiency.
With a higher quenching efficiency one can also improve the imaging resolution by using a doughnut pulsed laser beam at a lower wavelength, since PL quenching was also observed under 532~nm wavelength pulsed laser beam excitation of nanodiamond~\cite{Plakhotnik:2011fy}. The superposition in space and time of a high energy doughnut-shape 532~nm pulsed beam and a low energy Airy-shape beam at a slightly higher wavelength (to be able to reject the remaining 532~nm intense beam) should result in a smaller nanodiamond PL spot.

Finally, other applications such as fast switching of NV color center-based single-photon source and tunable photonic bandgap device based on diamond photonic crystal, could be also envisioned using the control IR pulsed beam.

\section*{Acknowledgements}

This work was supported by the European Commission through EQUIND (FP6 project number IST-034368) and NEDQIT (ERANET Nano-Sci) projects, by the Agence Nationale de la Recherche (France) through the project grant ANR-06-NANO-041, and by ``Triangle de la Physique'' contract B-DIAMANT. We gratefully thank D. Garrot and Xuan Loc Le for useful discussions and we acknowledge F. Druon and D. Papadopoulos for their loan of the 1064~nm pulsed laser source.

\section*{Appendix A. Heat transfer characteristic time estimates}

We will consider that temperature attained by the diamond nanocrystal after its heating by the IR laser is $\approx 600$~K, and will calculate the relaxation time for this temperature to decrease to 450~K, which is the one at which the quantum yield is identical to the room temperature one~\cite{Toyli:2012gl}). 
If we do not consider heat transfer by the conduction from the nanodiamond to the coverglass substrate the two remaining relaxation mechanisms are convection and radiation, that we will consider independently. 
If only radiation is at play, the change of nanocrystal temperature $dT$ during the time interval $dt$ is $\rho c_P {\cal V} dT=-\varepsilon \sigma {\cal S} T^4 dt$ where $\rho$ is the diamond volumic mass, $c_P$ is the heat capacity of diamond per unit of mass at constant pressure, ${\cal V}$ and ${\cal S}$ are respectively the volume and the surface of the nanocrystal, $\sigma$ is the Stefan's constant, and $\varepsilon$ is nanodiamond emissivity. According to Ref.\cite{Victor:1962dh}, $c_P\approx 0.5$~kJ.kg$^{-1}$.K$^{-1}$ in 273 to 1073~K temperature range. Considering also $\rho=3.5$~kg.m$^{-3}$, and a nanodiamond of spherical geometry with a radius of 15~nm, we estimated by integrating the above equation, that the temperature relaxation time to go from 600~K to 450~K is $\tau_{\rm rad}\approx 0.4$~ms, assuming $\varepsilon=1$, which is the largely overestimated because pure diamond has no emissivity. However due to the presence of numerous defect in nanodiamonds, mostly at their surface, the emissivity is expected to be non zero. The estimated $\tau_{\rm rad}\approx 0.4$~ms is therefore a lower bound.

Similarly, the purely convective relaxation obeys $\rho c_P {\cal V} dT=-h{\cal{S}}(T_{\rm air}-T)dt$, where $h$ is heat transfer coefficient of air approximately equal to 2~\cite{bejan1984convection}, and $T_{\rm air}$ the ambient air temperature. From this equation, we infer the convective relaxation $\tau_{\rm conv}\approx 3$~ms.

\section*{Appendix B. Effect of 1064~nm pulses on the photoluminescence of NV centers in bulk diamond crystal}

\newcounter{figappendix}
\setcounter{figappendix}{1}
\renewcommand{\thefigure}{B\arabic{figappendix}}

\begin{figure}[h!]
\centerline{\resizebox{\columnwidth}{!}{
\includegraphics{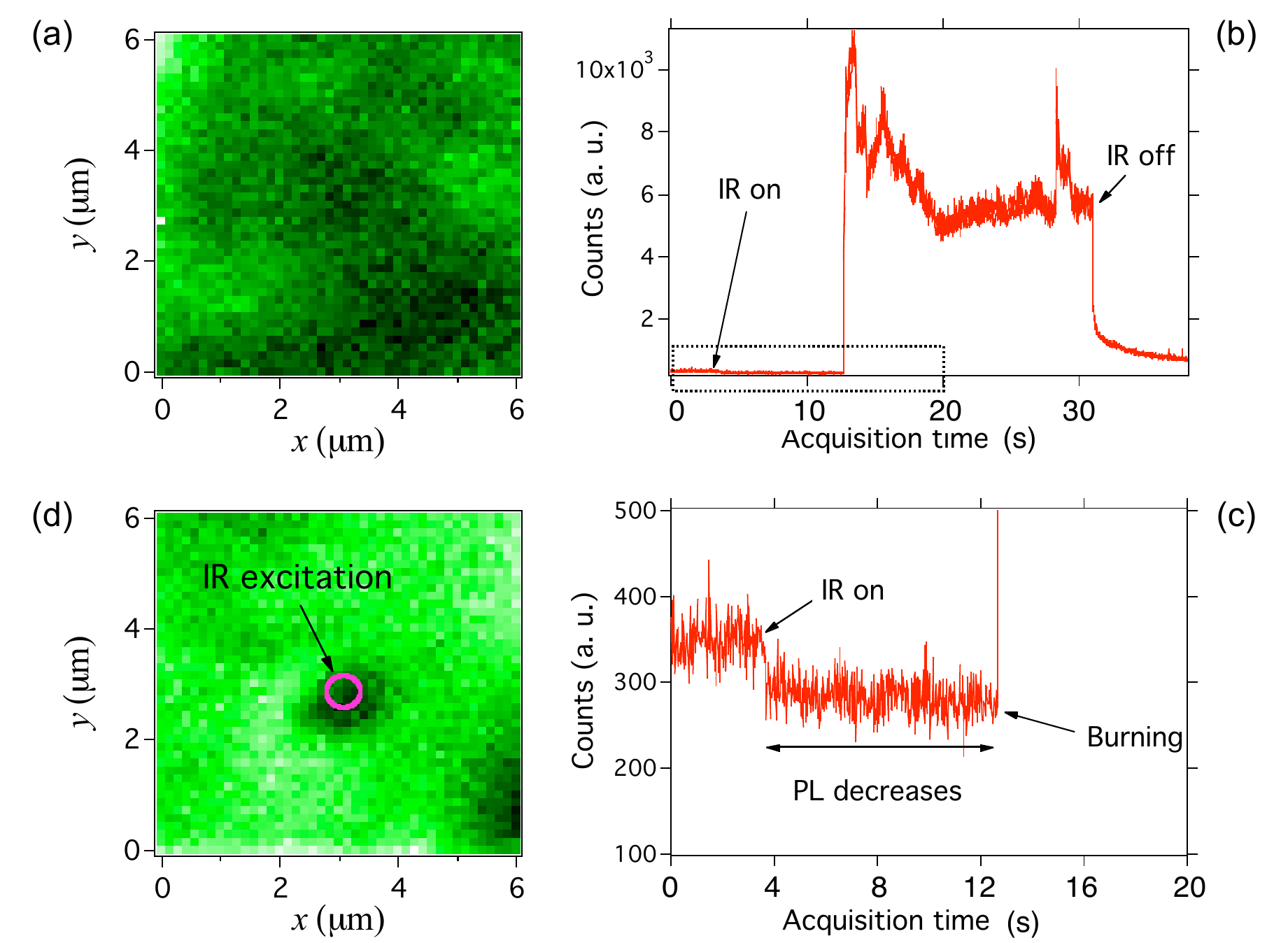}}}
\caption{1064~nm pulsed excitation effect on the photoluminescence of NV color centers in a bulk diamond crystal. 
(a) Photoluminescence rasterscan obtained with the 532~nm excitation beam alone, at 0.18~mW mean power. 
(b) Photoluminescence intensity timetrace obtained when the 1064~nm beam is added at time 4~s (see zoom in (c)) with 30~mW mean power. A bright light emission is observed starting from 13~s, and lasting until the IR is switched off. This emission is associated to the burning of diamond is hieratic and does not disappear instantaneously when the IR beam is switched off, maybe due to continuing burning. 
(c) Zoom of the photoluminescence intensity timetrace obtained at the beginning of the 1064~nm illumination. The photoluminescence signal slightly decreases by about 20\% when the IR beam is switched on about 4~s after the start of the monitoring, like in the case of nanodiamond, but the signal becomes suddenly very large at time 13~s due to diamond starting to burn.
(d) Photoluminescence rasterscan of the sample after exposure of the middle part of the scan to the IR beam. A black spot is observed at focusing point of the IR excitation beam.}
\label{fig:bulk}
\end{figure}

To evaluate the role of thermal effects in the PL quenching, we replaced the nanodiamonds by a bulk diamond sample containing NV$^-$ color centers, at a density of about 200 color centers per $\mu$m$^3$, deduced from the PL intensity (figure~\ref{fig:bulk}(a)) compared to the one of a single emitter.
Diamond possesses one of the highest thermal conductivity among solids. Therefore any heat dissipated after the IR pulsed illumination will diffuse very quickly in the whole sample, and will not be accompanied by a significant increase of the local temperature, contrary to the case of diamond nanocrystal. 

We focused the IR beam at the air-diamond interface surface, which results in an increase of the illumination intensity. After about 13~s of IR illumination, we observed a bright broadband light emission that we attribute to the burning of the diamond at the focus spot (figure~\ref{fig:bulk}(b)) consecutive to optical breakdown~\cite{Joglekar:2004cr}. This burning is observed at lower IR illumination power than in the case of nanodiamonds, because in the case of the bulk sample the whole incident laser power interacts with the diamond matrix. In the case of nanodiamond, only a fraction of the incident beam is absorbed due to its size smaller than the laser spot. 
For power larger than 40~mW the burning takes place at any position of the IR focus onto the sample surface. When the IR power is decreased to 20-30~mW, the effect becomes sensitive the focusing position, which may be due to inhomogeneous surface composition (presence of amorphous carbon for example).

Figure~\ref{fig:bulk}(a) displays a diamond bulk sample scanned with only the green excitation beam. We then  illuminated only the central part of the sample by the IR beam at 30~mW mean power (corresponding to an intensity of 18~MW/cm$^2$), and recorded simultaneously the PL intensity vs time (Figure~\ref{fig:bulk}(b)). At this IR power, the burning is not immediately observed. 
In fact, a decrease of photoluminescence is first observed as for fluorescent nanodiamonds (zoom shown in figure~\ref{fig:bulk}(c)) followed by the burning (figure~\ref{fig:bulk}(b)). A PL rasterscan carried out after the IR illumination shows a decrease of the PL intensity in the region exposed to the IR beam (figure~\ref{fig:bulk}(c)). 
When the IR beam is switched off at time $\approx 32$~s, the white light emission accompanying the diamond burning slowly stops (figure~\ref{fig:bulk}(b)).

These observations confirm the relation between the two effects: the NV center PL quenching and the burning of diamond, as a consequence of increasing IR illumination and optical breakdown. It is worth to note that the burning of diamond by high power near infrared pulsed illumination is used for diamond micromachining~\cite{Sudheer:2006uy,Zalloum:2010uv}.

\section*{References}

\providecommand{\newblock}{}
\begin{thebibliography}{10}
\expandafter\ifx\csname url\endcsname\relax
  \def\url#1{{\tt #1}}\fi
\expandafter\ifx\csname urlprefix\endcsname\relax\def\urlprefix{URL }\fi
\providecommand{\eprint}[2][]{\url{#2}}

\bibitem{Beveratos:2001ht}
Beveratos A, Brouri R, Gacoin T, Poizat J~P and Grangier P 2001 {Nonclassical
  radiation from diamond nanocrystals} {\em Phys. Rev. A\/} {\bf 64}

\bibitem{Jelezko:2004jn}
Jelezko F, Gaebel T, Popa I, Gruber A and Wrachtrup J 2004 {Observation of
  Coherent Oscillations in a Single Electron Spin} {\em Phys. Rev. Lett.\/}
  {\bf 92}

\bibitem{Wrachtrup:2011uz}
Wrachtrup J and Jelezko F 2011 {Focus on Diamond-Based Photonics and
  Spintronics} {\em New J. Phys.\/} {\bf 13}

\bibitem{Balasubramanian:2008ga}
Balasubramanian G, Chan I~Y, Kolesov R, Al-Hmoud M, Tisler J, Shin C, Kim C,
  Wojcik A, Hemmer P~R, Krueger A, Hanke T, Leitenstorfer A, Bratschitsch R,
  Jelezko F and Wrachtrup J 2008 {Nanoscale imaging magnetometry with diamond
  spins under ambient conditions} {\em Nature\/} {\bf 455} 648--651

\bibitem{Rondin:2012ku}
Rondin L, Tetienne J~P, Spinicelli P, Dal~Savio C, Karrai K, Dantelle G,
  Thiaville A, Rohart S, Roch J~F and Jacques V 2012 {Nanoscale magnetic field
  mapping with a single spin scanning probe magnetometer} {\em Appl. Phys.
  Lett.\/} {\bf 100} 153118

\bibitem{Maletinsky:2012ge}
Maletinsky P, Hong S, Grinolds M~S, Hausmann B, Lukin M~D, Walsworth R~L,
  Loncar M and Yacoby A 2012 {A robust scanning diamond sensor for nanoscale
  imaging with single nitrogen-vacancy centres} {\em Nature Nanotechnology\/}
  {\bf 7} 320--324

\bibitem{Chang:2008ia}
Chang Y~R, Lee H~Y, Chen K, Chang C~C, Tsai D~S, Fu C~C, Lim T~S, Tzeng Y~K,
  Fang C~Y, Han C~C, Chang H~C and Fann W 2008 {Mass production and dynamic
  imaging of fluorescent nanodiamonds} {\em Nature Nanotechnology\/} {\bf 3}
  284--288

\bibitem{Faklaris:2008bm}
Faklaris O, Garrot D, Joshi V, Druon F, Boudou J~P, Sauvage T, Georges P, Curmi
  P~A and Treussart F 2008 {Detection of Single Photoluminescent Diamond
  Nanoparticles in Cells and Study of the Internalization Pathway} {\em
  Small\/} {\bf 4} 2236--2239

\bibitem{Fang:2011gm}
Fang C~Y, V V, Cheng C~A, Yeh S~H, Chang C~F, Li C~L and Chang H~C 2011 {The
  Exocytosis of Fluorescent Nanodiamond and Its Use as a Long-Term Cell
  Tracker} {\em Small\/}  n/a--n/a

\bibitem{Rittweger:2009cc}
Rittweger E, Han K~Y, Irvine S~E, Eggeling C and Hell S~W 2009 {STED microscopy
  reveals crystal colour centres with nanometric resolution} {\em Nature
  Photon\/} {\bf 3} 144--147

\bibitem{ceggeli:2009wea}
Han K~Y, Willig K~I, Rittweger E, Jelezko F, Eggeling C and Hell S~W 2009
  {Three-Dimensional Stimulated Emission Depletion Microscopy of
  Nitrogen-Vacancy Centers} {\em Nano Lett.\/} {\bf 9} 3323--3329

\bibitem{Tzeng:2011do}
Tzeng Y~K, Faklaris O, Chang B~M, Kuo Y, Hsu J~H and Chang H~C 2011
  {super-resolution Imaging of Albumin-Conjugated Fluorescent Nanodiamonds in
  Cells by Stimulated Emission Depletion} {\em Angew. Chem. Int. Ed.\/} {\bf
  50} 2262

\bibitem{Maurer:2010js} Maurer P C, Maze J R, Stanwix P L, Jiang L, Gorshkov A V, Zibrov A A, Harke B, Hodges J S, Zibrov A S, Yacoby A, Twitchen D, Hell S W, Walsworth R L and Lukin M D 2010 Far-field optical imaging and manipulation of individual spins with nanoscale resolution {\em Nature Physics\/} {\bf 6} 912Ð-918

\bibitem{Plakhotnik:2010cg}
Plakhotnik T and Gruber D 2010 {Luminescence of nitrogen-vacancy centers in
  nanodiamonds at temperatures between 300 and 700 K: perspectives on
  nanothermometry} {\em Phys. Chem. Chem. Phys.\/} {\bf 12} 9751

\bibitem{Toyli:2012gl}
Toyli D, Christle D, Alkauskas A, Buckley B, Van~de Walle C and Awschalom D
  2012 {Measurement and Control of Single Nitrogen-Vacancy Center Spins above
  600 K} {\em Phys. Rev. X\/} {\bf 2} 031001

\bibitem{Plakhotnik:2011fy}
Plakhotnik T and Chapman R 2011 {Nitrogen-vacancy centers in nano-diamond
  reversibly decrease the luminescence quantum yield under strong pulsed-laser
  irradiation} {\em New J. Phys.\/} {\bf 13} 045001

\bibitem{g1994luminescent}
Blasse G and Grabmaier B~C 1994 {\em {Luminescent materials}\/}
  (Springer-Verlag)

\bibitem{Rittweger:2009gv}
Rittweger E, Wildanger D and Hell S~W 2009 {Far-field fluorescence nanoscopy of
  diamond color centers by ground state depletion} {\em Europhys. Lett.\/} {\bf
  86} 14001

\bibitem{Papadopoulos:2003hv}
Papadopoulos D~N, Forget S, Delaigue M, Druon F, Balembois F and Georges P 2003
  {Passively mode-locked diode-pumped Nd:YVO{\_}4 oscillator operating at an
  ultralow repetition rate} {\em Opt. Lett.\/} {\bf 28} 1838--1840

\bibitem{Rondin:2010dn} Rondin L, Dantelle G, Slablab A, Grosshans F, Treussart F, Bergonzo P, Perruchas S, Gacoin T, Chaigneau M, Chang H-C, Jacques V and Roch J F 2010 {Surface-induced charge state conversion of nitrogen-vacancy defects in nanodiamonds} {\em Phys. Rev. B \/} {\bf 82} 115449

\bibitem{Lai:2009gm}
Lai N~D, Zheng D, Jelezko F, Treussart F and Roch J~F 2009 {Influence of a
  static magnetic field on the photoluminescence of an ensemble of
  nitrogen-vacancy color centers in a diamond single-crystal} {\em Appl. Phys.
  Lett.\/} {\bf 95} 133101

\bibitem{Wee:2007dv}
Wee T~L, Tzeng Y~K, Han C~C, Chang H~C, Fann W, Hsu J~H, Chen K~M and Yu Y~C
  2007 {Two-photon Excited Fluorescence of Nitrogen-Vacancy Centers in
  Proton-Irradiated Type Ib Diamond} {\em J. Phys. Chem. A\/} {\bf 111}
  9379--9386

\bibitem{Xu:2002td} Xu~N~S, Chen~J and Deng~S~Z 2002 {Effect of heat treatment on the properties of nano-diamond under oxygen and argon ambient} {\em Diam. Relat. Mater. /} {\bf 11} 249Ð56

\bibitem{Rogers:2008ip}
Rogers L~J, Armstrong S, Sellars M~J and Manson N~B 2008 {Infrared emission of
  the NV centre in diamond: Zeeman and uniaxial stress studies} {\em New J.
  Phys.\/} {\bf 10} 103024

\bibitem{Glinka:1999efa}
Glinka Y~D, Lin K~W and Lin S~H 1999 {Multiphoton-excited luminescence from
  diamond nanoparticles and an evolution to emission accompanying the laser
  vaporization process} {\em Appl. Phys. Lett.\/} {\bf 74} 236

\bibitem{Zalloum:2010uv}
Zalloum O~H~Y, Parrish M, Terekhov A and Hofmeister W 2010 {On femtosecond
  micromachining of HPHT single-crystal diamond with direct laser writing using
  tight focusing.} {\em Opt. Express\/} {\bf 18} 13122--13135

\bibitem{Joglekar:2004cr}
Joglekar A~P 2004 {Optics at critical intensity: Applications to nanomorphing}
  {\em Proc. Natl. Acad. Sci. U.S.A.\/} {\bf 101} 5856--5861

\bibitem{Youngworth:2000hf}
Youngworth K and Brown T 2000 {Focusing of high numerical aperture
  cylindrical-vector beams} {\em Opt. Express\/} {\bf 7} 77--87

\bibitem{Greffet:2011tb}
Greffet J~J, Hugonin J~P, Besbes M, Lai N~D, Treussart F and Roch J~F 2011
  {Diamond particles as nanoantennas for nitrogen-vacancy color centers} {\em
  arXiv\/} {\bf 1107.0502v1}

\bibitem{Sudheer:2006uy}
Sudheer S~K, Pillai V~P~M and Nayar V~U 2006 {Diode pumped Q-switched Nd: YAG
  laser at 1064 nm with nearly diffraction limited output beam for precise
  micromachining of natural diamond for micro-electro-mechanical systems (MEMS)
  applications} {\em Journal of Optoelectronics and Advanced Materials\/} {\bf
  8} 363

\bibitem{Victor:1962dh}
Victor A~C 1962 {Heat Capacity of Diamond at High Temperatures} {\em J. Chem.
  Phys.\/} {\bf 36} 1903
  
  \bibitem{bejan1984convection}
Bejan A 1984 {\em Convection heat transfer\/} Wiley-Interscience publication
  (Wiley)
  
  \bibitem{Loo:2005iu} Loo C, Lowery A, Halas N, West J and Drezek R 2005 Immunotargeted Nanoshells for Integrated Cancer Imaging and Therapy {\em Nano Lett. \/} {\bf 5} 709Ð11
  
  \bibitem{Chang:2008hj} Chang C-C, Chen P-H, Chu H-L, Lee T-C, Chou C-C, Chao J-I, Su C-Y, Chen J S, Tsai J-S, Tsai C-M, Ho Y-P, Sun K W, Cheng C-L and Chen F-R 2008  Laser induced popcornlike conformational transition of nanodiamond as a nanoknif {\em Appl. Phys. Lett.\/} {\bf 93} 033905
\endbib

\end{document}